\documentclass[preprint,12pt,authoryear]{elsarticle}

\usepackage{amssymb}
 \usepackage{amsthm}
 \usepackage{amsmath}
 \usepackage{lineno}
  \usepackage{natbib}
 \theoremstyle{plain}
 \newdefinition{definition}{Definition}
\newtheorem{theorem}{Theorem}
 
  \newtheorem{corollary}[theorem]{Corollary}
 \newdefinition{remark}{Remark}
 \newdefinition{property}{Property}

\journal{arXiv}

\begin{document}

\begin{frontmatter}



\title{Optimal Designs for Copula Models}


\author{E. PERRONE  and W.G. M\"ULLER}

\address{Department of Applied Statistics, Johannes Kepler University Linz,\\ 4040 Linz, Austria}

\begin{abstract}
Copula modelling has in the past decade become a standard tool in many areas of applied statistics. However, a largely neglected aspect concerns the design of related experiments. Particularly the issue of whether the estimation of copula parameters can be enhanced by optimizing experimental conditions and how robust all the parameter estimates for the model are with respect to the type of copula employed. In this paper an
equivalence theorem for (bivariate) copula models is provided that allows formulation of efficient design algorithms and quick checks of whether designs are optimal or at least efficient. Some examples illustrate that in practical situations considerable gains in design efficiency can be achieved. A natural comparison between different copula models with respect to design efficiency is provided as well.

\end{abstract}


\end{frontmatter}


\section{Introduction}

Due to their flexibility in describing dependencies and the possibility of separating marginal and joint effects copula models have become a popular device for coping with multivariate data. in many areas of applied statistics eg. for insurances \citep{valdez_98}, econometrics \citep{trivedi+z_06}, medicine \citep{Nikoloulopoulos+k_08}, marketing \citep{danaher+s_10}, spatial extreme events (\citealp{wadsworth+t_12}), time series analysis \citep{patton_12}, even sports \citep{mchale+s_11} and particularly in finance (\citealp{cherubini+al_04}).

The concept of copulas, however, has only been rarely employed in experimental design with notable exceptions of spatial design in \cite{li+al_11} and \cite{pilz+al_12}, and sequential trials in \cite{schmidt+al_14}. The design question for copula parameter estimation has to our knowledge just been raised in \cite{denman+al_11}, where a brute-force simulated annealing optimization was employed for the solution of a specific problem.
By this paper we provide the necessary theory for fully embedding the situation into optimal design theory. Particularly we provide a Kiefer-Wolfowitz type equivalence theorem \citep{kiefer+w_60} in Section 3 as a basis for a substantial analysis of the arising issues in the example sections.

To be more concrete, let us consider a vector $\mathbf{x}^T = (x_1, \ldots, x_r) \in \mathcal{X}$ of control variables, where
$\mathcal{X} \subset \Re^r$ is a compact set. The results of the observations and of the expectations in a regression experiments are the vectors:
\[ \mathbf{y}(\mathbf{x}) = (y_1(\mathbf{x}), , \ldots, y_m(\mathbf{x})), \]
\[\mathbf{E}[\mathbf{Y}(x)] = \mathbf{E}[(Y_1, \ldots, Y_m)] = \mathbf{\eta}(\mathbf{x},\mathbf{\beta}) = (\eta_1(\mathbf{x},\mathbf{\beta}), \ldots, \eta_m(\mathbf{x},\mathbf{\beta})),\]
where $\mathbf{\beta}=(\beta_1, \ldots,\beta_k)$ is a certain unknown (trend) parameter vector to be estimated and $\eta_i (i = 1, \ldots,n)$ are known functions. In the remainder of the paper we will focus on the case $m = 2$, but generalizations of our results are possible.

Let us call $F_{Y_i}(y_i(\mathbf{x}, \mathbf{\beta}))$ the margins of each $Y_i$ for all $i=1,\ldots,m$ and $c_{\mathbf{Y}}(\mathbf{y}(\mathbf{x}, \mathbf{\beta}), \mathbf{\alpha})$ the joint probability density function of the random vector $\mathbf{Y}$, where $\mathbf{\alpha} = (\alpha_1,\ldots, \alpha_l)$ are unknown (copula) parameters.

\begin{definition}
\label{Def:Copula}
Let $\mathcal{I} = [0,1]$. A \emph{two-dimensional copula} (or \emph{2-copula}) is a bivariate function $C: \mathcal{I}\times\mathcal{I} \longrightarrow \mathcal{I}$ with the following properties:
\begin{enumerate}
\item	for every $u_1$, $u_2 \in \mathcal{I}$
		\begin{equation}
		C(u_1,0) = 0,  \; C(u_1, 1) = u_1,  \; C(0,u_2) = 0,  \; C(1,u_2) = u_2;
		\end{equation}
\item	for every $u_1$, $u_2$, $u_3$, $u_4 \in \mathcal{I}$ such that $u_1 \leq u_3$ and $u_2 \leq u_4$,
		\begin{equation*}
		C(u_3,u_4) - C(u_3,u_2) - C(u_1,u_4) + C(u_1,u_2) \geq 0.
		\end{equation*}
				
\end{enumerate}
\end{definition}

Now let ${\mathbf{F}}_{{Y_1}{Y_2}}$ be a joint distribution function with marginals $F_{Y_1}$ and $F_{Y_2}$.
According to Sklar's theorem \citep{Sklar_59} there exists then a 2-copula $C$ such that
\begin{equation}\label{Eq:S}
\mathbf{F}_{Y_1Y_2} (y_1,y_2) = C(F_{Y_1}(y_1), F_{Y_2}(y_2))
\end{equation}
for all reals $y_1$, $y_2$.
If $F_{Y_1}$ and $F_{Y_2}$ are continuous, then $C$ is unique; otherwise, $C$ is uniquely defined on $\text{Ran}(F_{Y_1})\times \text{Ran}(F_{Y_2})$.
Conversely, if $C$ is a 2-copula and $F_{Y_1}$ and $F_{Y_2}$ are distribution functions, then the function $F_{Y_1Y_2}$ given by (\ref{Eq:S}) is a joint distribution with marginals $F_{Y_1}$ and $F_{Y_2}$.

\section{Design issues}

We need to quantify the amount of information on both (trend and copula) sets of parameters $\alpha$ and $\beta$ respectively from the regression experiment embodied in the Fisher information matrix, which for a single information is a $(k + l) \times (k + l)$ matrix defined as
\begin{equation}
m(\mathbf{x}, \beta,\alpha) = \left(
\begin{array}{cc}
m_{\mathbf{\beta\beta}}(\mathbf{x}) &  m_{\mathbf{\beta\alpha}}(\mathbf{x}) \\
m_{\mathbf{\beta\alpha}}^T(\mathbf{x}) & m_{\mathbf{\alpha\alpha}}(\mathbf{x})
\end{array} \right)
\end{equation}
where the submatrix $m_{\mathbf{\beta\beta}}(\mathbf{x})$ is the $( k \times k)$ matrix with the $(i,j)$th element defined as
\begin{equation}\label{Eq:FIM}
 \mathbf{E} \left(  - \dfrac{\partial^2}{\partial \beta_i \partial \beta_j} \log[c_{\mathbf{Y}}(\mathbf{y}(\mathbf{x}, \mathbf{\beta}), \mathbf{\alpha})] \right)
 \end{equation}
and the submatrices $m_{\beta\alpha}(\mathbf{x})$ $( k \times l)$  and $m_{\alpha\alpha}(\mathbf{x})$ $( l \times l)$ are defined accordingly.
Here we model the dependence between $Y_1$ and $Y_2$ with a copula function $C_{\alpha}(F_{Y_1}(y_1(\mathbf{x}, \mathbf{\beta})), F_{Y_2}(y_2(\mathbf{x}, \mathbf{\beta})))$ and find the density of that copula from
\[ c_{\mathbf{Y}}(\mathbf{y}(\mathbf{x}, \mathbf{\beta}), \mathbf{\alpha}) = \dfrac{\partial^2}{\partial y_1 \partial y_2} C_{\alpha}(F_{Y_1}(y_1(\mathbf{x}, \mathbf{\beta})), F_{Y_2}(y_2(\mathbf{x}, \mathbf{\beta}))).\]

\begin{definition}
\label{Def:DesMeasure}
A probability distribution function $\xi$ on the actual design
space $\Xi$ , which is the class of all the probability distributions on the Borel set $\mathcal{X}$, is called a \emph{design measure}.
\end{definition}

The Information Matrix on a general design measure is $M(\xi, \beta, \alpha) = E(m(\tilde{x}, \beta,\alpha))$ where $\tilde{x}$ is a random vector with distribution $\xi$. So for $r$ independent observations at $x_1,\ldots,x_r$, the corresponding Information matrix is
\[\mathbf{M}(\xi, \beta, \alpha) = \sum\limits_{i=1}^r w_i m(x_i,\beta,\alpha), \sum\limits_{i=1}^r w_i = 1,
\xi = \left \{
\begin{array}{cccc}
x_1 &  \ldots & x_n \\
w_1 &  \ldots & w_n
\end{array}
\right \},
\]
and the aim of approximate optimal design theory is concerned with finding $\xi^*(\beta, \alpha)$ such that it maximizes some scalar function $\phi(M(\xi, \beta, \alpha))$, the so-called design criterion. In the following we will consider only \emph{D-optimality}, i.e. the criterion $\phi (M) = \log \det M $, if $M$ is non singular. There exist several well written monographs on optimal design theory and its application, but in this paper we follow mainly the style and notation of \cite{silvey_80}.

\section{Equivalence theory}

The cornerstone of a theoretical investigation into optimal design is usually the formulation of  a Kiefer-Wolfowitz type equivalence relation, which is given in the following theorem. It is a generalized version of a theorem given without proof in \cite{heise+m_1996} and follows from a multivariate version of the basic theorem given in  \cite{silvey_80}, its full proof can be found in the supplementary material.

\begin{theorem}
\label{theorem1}
For a local parameter vector $(\bar{\beta}, \bar{\alpha})$, the following properties are equivalent:
\begin{itemize}
\item $\xi^*$ is D-optimal;
\item $\textnormal{ tr }[ M(\xi^*, \bar{\beta},\bar{\alpha})^{-1} m(x, \bar{\beta},\bar{\alpha})]\leq (k+l)$, $\forall x \in \mathcal{X}$;
\item $\xi^*$ minimize $\max\limits_{x \in \mathcal{X}}\textnormal{ tr }[M(\xi^*, \bar{\beta},\bar{\alpha})^{-1} m(x, \bar{\beta},\bar{\alpha})]$, over all $\xi \in \Xi$.\end{itemize}

\end{theorem}

This theorem allows us the use of standard design algorithms such as of the Fedorov-Wynn-type \citep{fedorov_71, wynn_70}. It also provides simple checks for D-optimality through the maxima of $$d(x,\xi^*)= \text{ tr }[ M(\xi^*, \bar{\beta},\bar{\alpha})^{-1} m(x, \bar{\beta},\bar{\alpha})],$$ which is usually called \emph{sensitivity function}.

\begin{definition}
For the comparison of two different designs define the ratio
\begin{equation}
 \left( \dfrac{|M(\xi, \beta, \alpha)|}{|M(\xi',\beta,\alpha)|}\right)^{1/(k+l)}
\end{equation}
where $(k+l)$ is the number of the model parameters, which is called \emph{D-Efficiency} of the design $\xi$ with respect to the design $\xi'$.
\end{definition}

Note that the resulting optimal designs will now depend not only upon the trend model structure, but also upon the chosen copula and through the induced nonlinearities potentially also on the unknown parameter values for $\alpha$ and $\beta$, which is why we are resorting to localized designs around the values $(\bar{\beta}, \bar{\alpha})$. A main question of course concerns whether ignorance or wrong guesses of copula function and/or parameters may lead to inefficiencies of the designs.

\section{Examples}
\subsection{Tools}
For that purpose let us here give the list of copulas used in our examples (for more details see eg. \citealp{Nelsen_06} or \citealp{durante+s_10}). We provide the copula function along with the so-called Kendalls $\tau$, which is a dependence measure that allows us to conveniently relate different copulas (for a definition and a more exhaustive comparison see \citealp{michiels+d_08}).
\begin{definition}
\begin{enumerate}
\item[]
\item \emph{Product Copula,} which represents the independence case.
\[C(u_1,u_2) = u_1 u_2,\] with $\tau = 0$.
\item[]
\item \emph{Gaussian Copula.}
\item[] \[{C}_{\alpha}(u_1,u_2) = \dfrac{1}{2 \Pi \sqrt{1- \alpha^{2}}} \times\]

\[\int_{-\infty}^{\Phi^{-1}(u_1)} \int_{-\infty}^{\Phi^{-1}(u_2)}
\exp \left(  - \dfrac{{z_1}^2 - 2 \alpha {z_1} {z_2} + {z_2}^2}{ 2 (1 - \alpha^{2})} \right)
 d z_1 d z_2, \]
with  $\alpha \in [-1,1]$ and $\tau = \frac{2}{\Pi} \arcsin(\alpha)$ .
\item[]
\item \emph{Farlie-Gumbel-Morgenstern (FGM).}
\item[] \[ {C}_{\alpha}(u_1,u_2) = u_1 u_2[1 + \alpha (1-u_1)(1-u_2)],\] with $\alpha \in [-1,1]$ and $\tau = \frac{2}{9} \alpha$.
\item[]   
\item  \emph{Clayton.}
\item[]  \[{C}_{\alpha}(u_1,u_2) =\big[  \max\big( u_1^{-\alpha} + u_2^{-\alpha} -1 ,\, 0\big) \big]^{-\frac{1}{\alpha}},\] 
    with $\alpha \in (0, +\infty)$ and $\tau = \frac{\alpha}{\alpha + 2}$.
\item[]
\item \emph{Frank.}
\item[] \[{C}_{\alpha}(u_1,u_2) =  - \frac{1}{\alpha} \ln\big( 1 + \dfrac{(e^{-\alpha u_1} - 1) (e^{-\alpha u_2} - 1) }{ e^{-\alpha} - 1}  \big),\] with $\alpha \in (-\infty,+\infty)$, and  $\tau = 1 - \frac{4}{\alpha}(1 - \frac{1}{\alpha}\int_0^{\alpha} \frac{t}{e^t -1} d t)$.\\
\item \emph{Gumbel.}
\item[] \[{C}_{\alpha}(u_1,u_2) =\exp \big( - \big[ ( -\ln u_1)^{\alpha} + (-\ln u_2)^{\alpha} \big]^{\frac{1}{\alpha}} \big),\] with $\alpha \in [1, +\infty)$ and $\tau = \frac{\alpha - 1}{\alpha}$.
\end{enumerate}
\end{definition}

\subsection{The linear case}

Let us first consider a simple example reported in \cite{fedorov_71}. For each design point $x$, we may observe an independent pair of random variables $Y_1$ and $Y_2$, such that
\[ E[Y_1(x)] = \beta_0 + \beta_1 x + \beta_2 x^2 ,\]
\[E[Y_2(x)] = \beta_3 x + \beta_4 x^3 + \beta_5 x^4, \qquad 0 \leq x \leq 1.\]

\begin{figure}
\begin{center}
\includegraphics[width=0.6\textwidth]{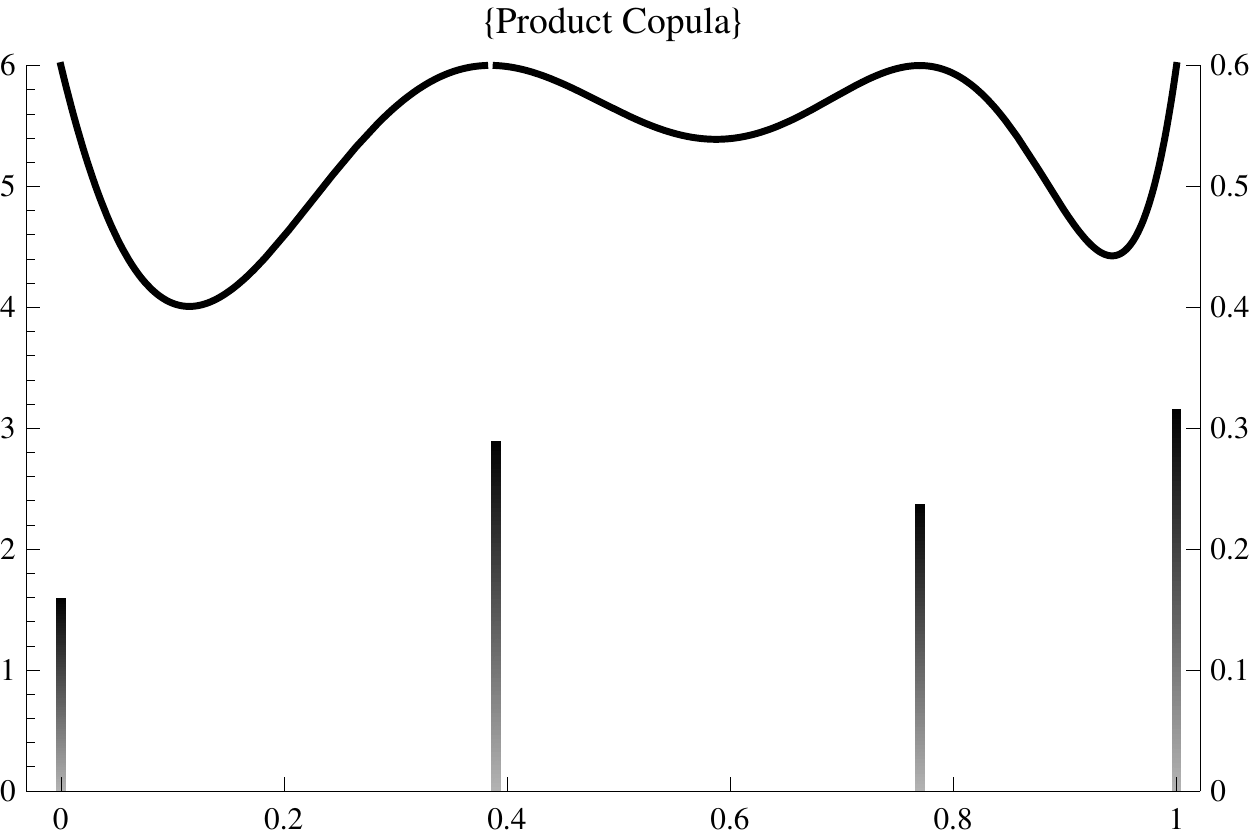}
\end{center}
\caption{Sensitivity function (left axis) and optimal design (right axis) for the Fedorov example.}
\label{figure1}
\end{figure}

This case is covered by Theorem 1 and employing the product copula and Gaussian margins. The optimal design which we have computed is given in Figure \ref{figure1} and is the same as reported in \cite{fedorov_71}, namely
\begin{equation} \label{xi*}
\xi^* =
\left( \begin{array}{c}
w_i \\
 x_i
\end{array}\right) =
\left( \begin{array}{cccc}
0.16 &  0.28 & 0.23 &  0.33 \\
 0 &  0.38 &  0.76 &  1.0
\end{array}\right).
\end{equation}

Let consider a more general case, when the joint distribution is described by a Gaussian copula and we thus allow the random variables $Y_1$ and $Y_2$ to be dependent. In this case the joint probability function of the random vector $\mathbf{Y}=(Y_1,Y_2)$ is simply
\begin{equation}
\begin{array}{cccc}
F_{\mathbf{Y}}(y_1,y_2) & = & C_{\alpha}(\Phi(y_1 - \eta_1(x,\beta)), \Phi(y_2 - \eta_2(x,\beta))) \\
& = & \Phi_2(y_1 - \eta_1(x,\beta),y_2 - \eta_2(x,\beta);\alpha)
\end{array}
\end{equation}
where $\Phi_2(\cdot,\cdot;\alpha)$ denotes the bivariate normal cumulative distribution function with correlation $\alpha \in (-1,1)$ and $\Phi$ denotes the cumulative distribution function of the standard normal distribution $N(0,1)$ (see \citealp{meyer_09}).

Our computations gave rise to the following
\begin{corollary}
For different values of $\alpha$ the optimal design is the same as for the independence case, which is the Gaussian case with $\alpha = 0$.
\end{corollary}
Note, that the sensitivity function now has a different scaling (with a maximum at $7$) as we have an additional copula parameter. This corollary, however, is hardly surprising as this fact coincides with the classic findings for the multivariate Gaussian distribution by \cite{krafft+s_92}.

But for a contrast consider now the Farlie-Gumbel-Morgenstern copula.
Following our approach, we must calculate the density of the function:
\[  {C}_{\alpha}(\Phi(Y_1(x;\mathbf{\beta})),\Phi(Y_2(x;\mathbf{\beta}))) = \]
\[\Phi(Y_1(x;\mathbf{\beta})) \Phi(Y_2(x;\mathbf{\beta}))[1 + \alpha (1-\Phi(Y_1(x;\mathbf{\beta})))(1-\Phi(Y_2(x;\mathbf{\beta})))],\]
which eventually leads to expressions like
\[\mathbf{E} \left(  - \dfrac{\partial^2}{\partial \beta_i \partial \beta_j} \log \left[ \dfrac{\partial^2}{\partial y_1 \partial y_2} {C}_{\alpha}(\Phi(Y_1(x;\mathbf{\beta})),\Phi(Y_2(x;\mathbf{\beta}))) \right] \right)\]
for the information matrix. These integrals are not analytically solvable, but we can evaluate them numerically and we can use the algorithm in order to find the optimum designs.

\begin{table}
\begin{center}{
\begin{tabular}{|r||r|r|r|r|r|r|}
\hline
 &\multicolumn{2}{|c|} { \textbf{FGM}}& \multicolumn{2}{|c|} {\textbf{Clayton}} & \multicolumn{2}{|c|} { \textbf{Frank}} \\
\hline
 \multicolumn{1}{|c||}{$\tau$} &  \multicolumn{1}{|c|}{$\alpha$} & D-eff &  \multicolumn{1}{|c|}{$\alpha$} & D-eff &  \multicolumn{1}{|c|}{$\alpha$} & D-eff\\
\hline
 $-0.15$ & $-0.67$ & $\mathbf{17.37}$ & n.d.& \multicolumn{1}{|c|}{-} & -1.37 & $\mathbf{0.10}$\\
\hline
 $-0.10$ & $-0.45$ & $\mathbf{0.23}$ & n.d. &\multicolumn{1}{|c|}{-} & -0.90 & $\mathbf{0.10}$\\
\hline
  $-0.05$ & $-0.22$ & $\mathbf{0.59}$ & n.d.& \multicolumn{1}{|c|}{-} & -0.45 & $\mathbf{0.10}$\\
\hline
 $0.05$ & $0.22$ & $\mathbf{0.68}$ & 0.10 & $\mathbf{0.16}$ & 0.45 & $\mathbf{0.10}$\\
\hline
 $0.10$ & $0.45$ & $\mathbf{0.39}$ & 0.22 & $\mathbf{0.13}$ & 0.90 & $\mathbf{0.10}$\\
\hline
 $0.15$ & $0.67$ & $\mathbf{10.18}$ & 0.35 & $\mathbf{0.34}$ & 1.37 & $\mathbf{0.10}$\\
\hline
 $0.35$ & n.d. & \multicolumn{1}{|c|}{-} & 1.08 & $\mathbf{0.11}$ & 3.51 & $\mathbf{0.11}$ \\
\hline
 $0.75$ & n.d. & \multicolumn{1}{|c|}{-} & 6.00 & $\mathbf{0.27}$ & 14.13 & $\mathbf{0.16}$ \\
\hline\hline
\end{tabular}
}
\end{center}
\caption{Losses in D-efficiency (in bold) by ignoring the dependence in percent.}
\label{table1}
\end{table}

The results are subsumed in Table \ref{table1}, which displays the loss in D-efficiency that occurs by using the optimal design $\xi^*$ from (\ref{xi*}) compared to the respective optimal designs for various copula models and Kendall's $\tau$. It can be seen that these losses are generally quite small, except perhaps for extreme values of $\tau$ in the FGM model.

\subsection{A binary bivariate model}
Let us now do the same for a more elaborate case with potential application in clinical testing. We consider a bivariate binary response $(Y_{i1}, Y_{i2})$, $i=1, \ldots, n$ with four possible outcomes $\{ (0,0),(0,1),(1,0),(1,1)\}$ where $1$ usually represents a success and $0$ a failure (of eg. a drug treatment).
For a single observation denote the joint probabilities of $Y_1$ and $Y_2$ by $p_{y_1,y_2} = pr (Y_1 = y_1, Y_2 = y_2)$ for $(y_1,y_2 = 0,1)$.

Now, define
\begin{equation}
p_{11} = C_{\alpha} (\pi_1, \pi_2), \quad p_{10} = \pi_1 - p_{11},  \quad p_{01} = \pi_2 - p_{11} \text{,   } \quad
p_{00} = 1 - \pi_1 - \pi_2 + p_{11}.
\end{equation}
The complete log-likelihood for the bivariate binary model is then given by
\begin{equation}
l(\mathbf{\theta}; \mathbf{y}) = \sum_{i=1}^n w_i l_i(\mathbf{\theta}; \mathbf{y}), \quad \mathbf{\theta} = (\mathbf{\beta_1}, \mathbf{\beta_2}, \alpha),
\end{equation}
where $w_i$ are the design weights and the log-likelihood for a single observation is given by
\begin{equation}
\begin{array}{ccc}
l_i(\mathbf{\theta}; \mathbf{y}) & = &  y_1 \, y_2 \, \log p_{11} \, +  y_1 \, (1- y_2) \, \log p_{10} \, + \\
 & & \\
& & (1- y_1) y_2 \log p_{01} +  (1 - y_1) (1 - y_2) \log p_{00}.
\end{array}
\end{equation}

As shown in \cite{dragalin+f_06} the Fisher information matrix for a single observation can then be written as
\begin{equation}
\mathbf{M}(\mathbf{\theta}, \xi_i) =
{\dfrac{\partial \mathbf{p}}{\partial \mathbf{\theta}}}^T
\left( P^{-1}
+ \dfrac{1}{1 - p_{11} - p_{10} - p_{01} }
 \mathbf{e}\mathbf{e}^T \right)
\dfrac{\partial \mathbf{p}}{\partial \mathbf{\theta}},
\end{equation}
where $\mathbf{p} = (p_{11}, p_{10}, p_{01})$, $P = diag(\mathbf{p})$ and $\mathbf{e} = (1,1,1)^T$. Some useful formulae for calculating information matrices in copula models can also be found in \cite{schepsmeier+s_14}.

The following example has initially been proposed in \cite{denman+al_11}. They assumed marginal probabilities of success given by the models
\begin{equation}
\log \left( \dfrac{\pi_i}{1 - \pi_i} \right) = \beta_{i0} + \beta_{i1} x, \qquad i=1,2
\end{equation}
where $x \in [0,10]$ and the initial parameters were $\mathbf{\beta_1}=[ -1, 1]$ and $\mathbf{\beta_2}=[ -2, 0.5]$.\\
They also investigated the three different copulas Frank, Clayton and Gumbel in order to make comparisons between the resulting designs. Note that in their calculations they employed a brute-force simulated annealing algorithm and had no means for checking definitive optimality, which is now possible through the equivalence theorem (\ref{theorem1}) provided. Note that the correlation range is restricted for these three copulae chosen, but we are generally not dependent upon this choice (\citealp{demirtas_13}).

\begin{figure}
{\begin{tabular}{cc}
\includegraphics[width=0.4625\textwidth]{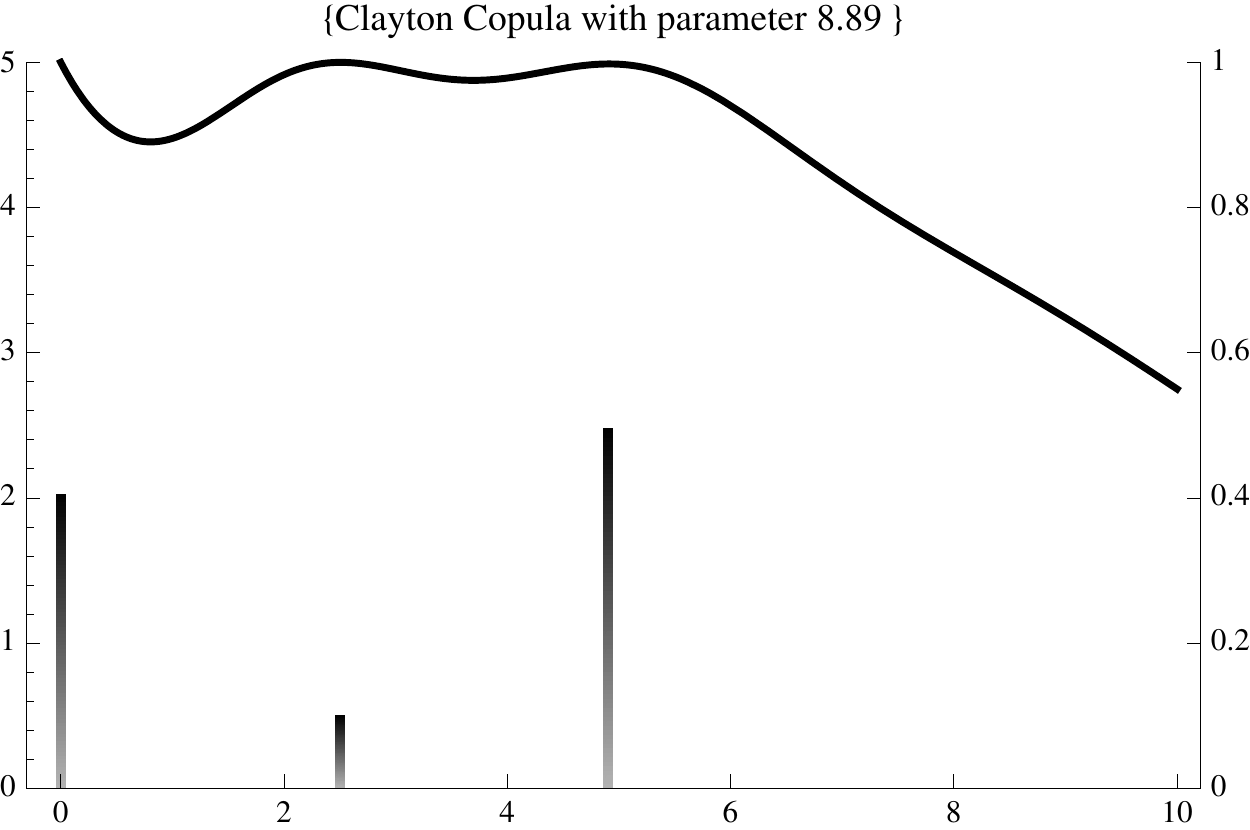}
&
\includegraphics[width=0.4625\textwidth]{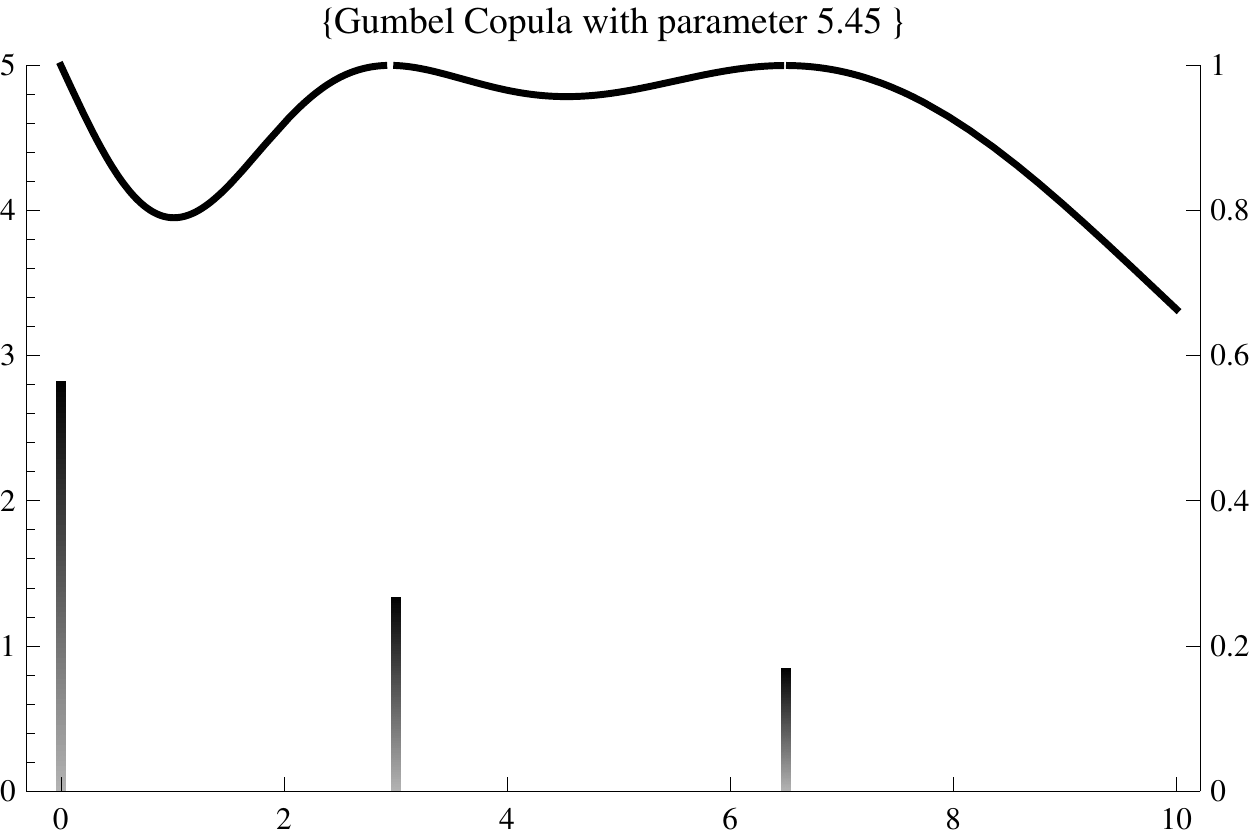}
\end{tabular}}\caption{The optimal designs and the sensitivity functions for the binary example (Clayton left, Gumbel right). The copula parameters chosen correspond to  Kendall's $\tau=0.816$.}
\label{figure2}
\end{figure}

So again by ignoring the dependence by not estimating the copula parameters, i.e. using just a four parameter model, for all copulas the same optimal design is found, which is given by
\[ \xi^*= \left(\begin{array}{c} w_i \\ x_i \end{array}\right) = \left(\begin{array}{ccc}
0.42 & 0.36 & 0.22 \\
>0 & 2.80 & 6.79
\end{array}\right).  \]
Using this design as a benchmark we note the losses in D-efficiency as reported in Table \ref{table2}. These losses are now stronger than in the linear case and seem to (at least for the Frank and Gumbel copula) grow with the dependence, as is intuitive. In Figure \ref{figure2} we display the designs and sensitivity functions for a representative case.

\cite{denman+al_11} also compared designs for various copula choices against each other in their Table 8. However, they have been using the same parameter values for these copulas, which does not seem sensible. We instead provide a comparison along the same Kendall's $\tau$ values in Table \ref{table2}, which naturally now shows much smaller discrepancies.


\begin{table}
\begin{center}{
\begin{tabular}{|r||r|r|r|r|r|r|}
\hline
  & \multicolumn{2}{|c|} { \textbf{Frank}}& \multicolumn{2}{|c|}{\textbf{Clayton}} & \multicolumn{2}{|c|}{\textbf{Gumbel}} \\
\hline
 \multicolumn{1}{|c||}{$\tau$} &  \multicolumn{1}{|c|}{$\alpha$} & D-eff &  \multicolumn{1}{|c|}{$\alpha$} & D-eff &  \multicolumn{1}{|c|}{$\alpha$} & D-eff\\
\hline
$0.11$ & 1.00 & $\mathbf{1.72}$ & 0.24 & $\mathbf{1.75}$ & 1.12 & $\mathbf{0.95}$\\
\hline
$0.45$ & 5.00 & $\mathbf{1.31}$ & 1.68 & $\mathbf{1.49}$ & 1.84 & $\mathbf{1.29}$\\
\hline
$0.66$ & 10.00& $\mathbf{1.87}$ & 3.98&  $\mathbf{0.71}$ & 3.00 & $\mathbf{2.31}$ \\
\hline
$0.76$ & 15.00 & $\mathbf{2.89}$ & 6.42 & $\mathbf{2.84}$ & 4.21 & $\mathbf{2.99}$\\
\hline
$0.82$ & 20.00 & $\mathbf{3.10}$ & 8.89 & $\mathbf{9.48}$ & 5.45 & $\mathbf{3.25}$ \\
\hline\hline
\end{tabular}
}
\end{center}
\caption{Losses in D-efficiency (in bold) by ignoring the dependence in percent.}
\label{table2}
\end{table}

\begin{table}
\begin{center}{
\begin{tabular}{|r||r|r|r|r|r|r|}
\hline
True Copula  & \multicolumn{2}{|c|} { \textbf{Frank}}& \multicolumn{2}{|c|}{\textbf{Clayton}} & \multicolumn{2}{|c|}{\textbf{Gumbel}} \\
\hline
Assumed Copula &  \multicolumn{1}{|c|}{Clayton} & \multicolumn{1}{|c|}{Gumbel} &  \multicolumn{1}{|c|}{Frank} & \multicolumn{1}{|c|}{Gumbel} &  \multicolumn{1}{|c|}{Frank} & \multicolumn{1}{|c|}{Clayton}\\
\hline
$\tau = 0.11$ & 2.24 & 0.67 & 1.99 & 2.70 & 0.82 & 2.75\\
\hline
$\tau = 0.45$ & 0.26 & 0.03 & 0.26 & 0.11 & 0.03 & 0.15\\
\hline
$\tau = 0.66$ & 1.09 & 0.11 & 1.04 &  1.28 & 0.14 & 1.57\\
\hline
$\tau = 0.76$ & 4.27 & 0.02 & 3.87 & 4.08 & 0.01 & 4.73\\
\hline
$\tau = 0.82$ & 8.24 & 0.01 & 10.91 & 10.96 & 0.01 & 8.43\\
\hline\hline
\end{tabular}
}
\end{center}
\caption{Losses in D-efficiency by comparing the true copula model with the assumed one for a fixed Kendall's tau value.}
\label{table3}
\end{table}

\section{Discussion}

Although the effects of ignoring the copula parameter seem to be rather mild judging by our examples, we expect stronger effects for some more non-symmetric copulae (see eg. \citealp{klement+m_06}), which are subject to our current research.

In general, our theory forms the basis to investigate further showcase examples from the literature, like e.g. in \cite{oakes+r_00}
or eventually treat mixed discrete/continuous type models like in \cite{deleon+w_10}. Particularly for the latter, but also quite generally the methods provided in this paper can thus be expected to be valuable for real applications from clinical trials, environmental sampling, industrial experiments, etc..

It is certainly of interest to extend the methods to models for which the copula parameters themselves are model-dependent such as in \cite{noh+al_13}, which we plan to do future research on.

\section*{Acknowledgement}
This was funded by the joint ANR/FWF project I833-N18: DESIRE. We thank M. Stehl\'ik, L. Pronzato, J. Rendas and E.P. Klement for fruitful discussions.

\appendix
\section{}
\subsection*{Equivalence Theorem}
Let us look at the design measure as a probability distribution function $\xi$ on the actual design
space $\Xi$ as opposed to the induced design space $\mathcal{X}$. Practically, $\Xi$ is the class of all the probability distributions on the Borel set $\mathcal{X}$ and is called \emph{design space}. For all the basics in what follows cf. \cite{silvey_80}.

For a given vector of parameters $(\beta, \alpha)$, let $\mathcal{M}_{(\beta, \alpha)}$ be the set of the information matrices generated as $\xi$ ranges over the class of all set of probability distribution on $\mathcal{X}$.

Then $\mathcal{M}_{(\beta, \alpha)}$ is the convex hull of $\{ m(x,\beta,\alpha): x \in \mathcal{X}\}$.

Let give now the definition of two derivatives that will play an important role in our theory.

\begin{definition}[G\^{a}teaux and Fr\'echet derivative]
\quad \\
Considering two elements $M_1$ and $M_2$ in $\mathcal{M}$, the \emph{G\^{a}teaux derivative of $\phi$} at $M_1$ in the direction of $M_2$ is:
\[ G_{\phi}(M_1, M_2) = \lim\limits_{\varepsilon \rightarrow 0^+} \frac{1}{\varepsilon}\{
\phi(M_1 + \varepsilon M_2) - \phi(M_1)\}, \]
the \emph{Fr\'echet derivative of $\phi$} at $M_1$ in the direction of $M_2$ is:
\[ F_{\phi}(M_1, M_2) = \lim\limits_{\varepsilon \rightarrow 0^+} \frac{1}{\varepsilon}\{
\phi\{(1- \varepsilon) M_1 + \varepsilon M_2\} - \phi(M_1)\}. \]
\end{definition}

The following are the properties of the derivatives that we defined before:
\begin{property}
The concavity of $\phi$ implies that
$$\frac{1}{\varepsilon} [\phi \{(1- \varepsilon) M_1 + \varepsilon M_2\} - \phi (M_1)]$$
is a non-increasing function of $\varepsilon$ in $0 < \varepsilon \leq 1$.
Hence when $\phi$ is concave, $F_{\phi} (M_1, M_2)$ exists if we allow the value $+ \infty$. \\

It is clear that if we put $\varepsilon =1$ in the previous equation, we obtain:
$ F_{\phi}(M_1, M_2) \geq \phi(M_2) + \phi(M_1).$ \\

According to the definitions of Fr\'echet and G\^{a}teaux derivatives, we can stress the following relationship between them:
$ F_{\phi}(M_1, M_2) = G_{\phi}(M_1, M_2 - M_1).$

Then, if we assume the differentiability of $\phi$ it is clear that
$$F_{\phi}(M_1, \sum a_iM_i) = \sum a_i F_{\phi}(M_1, M_i).$$

So, if $\tilde{M}$ is a random matrix, the following equivalence holds:
$$ E\{F_{\phi} (M_1 , \tilde{M}_1) \} = F_{\phi} \{M_1, E(\tilde{M_1})\} $$

\end{property}

\begin{theorem} \label{Teo:EqT}
Suppose to have a fixed parameters vector $(\bar{\beta}, \bar{\alpha})$, a concave function $\phi$ on $\mathcal{M}_{(\bar{\beta}, \bar{\alpha})}$ which is also differentiable at all points of $\mathcal{M}_{(\bar{\beta}, \bar{\alpha})}$ where $\phi(M) < - \infty$, so where a $\phi$ optimal measure exists.

Then the following are equivalent:
\begin{itemize}
 \item $\xi^*$ is $\phi$-optimal;
\item $F_{\phi} (M(\xi^*, \bar{\beta}, \bar{\alpha}), M(\xi,\bar{\beta}, \bar{\alpha} )) \leq 0 $, $\forall \xi \in \Xi$ ;
\item $F_{\phi} (M(\xi^*, \bar{\beta}, \bar{\alpha}), m(x,\bar{\beta}, \bar{\alpha} )) \leq 0 $, $\forall x \in \mathcal{X}$  ;
\item $\max\limits_{x \in \mathcal{X}} F_{\phi}(M(\xi^*, \bar{\beta}, \bar{\alpha}), m(x, \bar{\beta}, \bar{\alpha})) = \min\limits_{\xi \in \Xi} \max\limits_{x \in \mathcal{X}} F_{\phi}(M(\xi, \bar{\beta}, \bar{\alpha}), m(x, \bar{\beta}, \bar{\alpha})) $.
\end{itemize}

\begin{proof} Let us prove the theorem by double implications.

$(i) \Rightarrow (ii)$ \\

 $\xi^*$ is $\phi$-optimal.

This means that $\phi(M(\xi^*, \bar{\beta}, \bar{\alpha}))$ is maximal.

For the properties of the function $\phi$, the following relation holds:
\[\phi\{(1 - \varepsilon)M(\xi^*, \bar{\beta}, \bar{\alpha}) + \varepsilon M(\xi,\bar{\beta}, \bar{\alpha})\} - \phi\{M(\xi^*, \bar{\beta}, \bar{\alpha})\} \leq 0\]
for $\varepsilon \in [0,1]$ and all $\xi \in \Xi$.

For all the elements of $\mathcal{M}_{(\bar{\beta}, \bar{\alpha})}$ holds that
\[ (1 - \varepsilon)M(\xi^*, \bar{\beta}, \bar{\alpha}) + \varepsilon M(\xi,\bar{\beta}, \bar{\alpha}) = M\{(1 - \varepsilon)\xi^* + \varepsilon \xi \}\]
and this means, from the definition of the Fr\'echet derivative, that
\[F_{\phi}\{M(\xi^*, \bar{\beta}, \bar{\alpha}), M(\xi, \bar{\beta}, \bar{\alpha})\} \leq 0\]
for all $\xi \in \Xi$.\\

$(ii) \Rightarrow (iii)$ \\

Since $m(x,\bar{\beta}, \bar{\alpha})$ are elements of the convex hull $\mathcal{M}_{(\bar{\beta}, \bar{\alpha})}$, the condition $(iii)$ follows directly from the hypothesis. \\

$(iii) \Rightarrow (iv)$

Let us observe that if $\tilde{x}$ is a random vector with distribution $\xi$, the following equivalence is verified:
\[
\begin{array}{cccc}
 & E[F_{\phi}\{M(\xi, \bar{\beta}, \bar{\alpha}), m(\tilde{x}, \bar{\beta}, \bar{\alpha})\}] & =
\\  = &F_{\phi}\{M(\xi, \bar{\beta}, \bar{\alpha}), E[m(\tilde{x},\bar{\beta}, \bar{\alpha})] \}& = \\
 = & F_{\phi}\{ M(\xi,\bar{\beta}, \bar{\alpha}), M(\xi,\bar{\beta}, \bar{\alpha})\} & = & 0.
\end{array}
\]
So, it must be that:
\[
\max\limits_{x \in \mathcal{X}} F_{\phi}\{M(\xi, \bar{\beta}, \bar{\alpha}), m(x, \bar{\beta}, \bar{\alpha})\} \geq 0
\]
But, according to the hypothesis, we have that for the design $\xi^*$
\[
\max\limits_{x \in \mathcal{X}} F_{\phi}\{M(\xi^*, \bar{\beta}, \bar{\alpha}), m(x, \bar{\beta}, \bar{\alpha})\} \leq 0.
\]
Hence
\[
\begin{array}{cccc}
  \max\limits_{x \in \mathcal{X}} F_{\phi}\{M(\xi^*, \bar{\beta}, \bar{\alpha}), m(x, \bar{\beta}, \bar{\alpha})\} = 0  =& \\
 = \min\limits_{\xi} \max\limits_{x \in \mathcal{X}} F_{\phi}\{ M(\xi, \bar{\beta}, \bar{\alpha}), m(x, \bar{\beta}, \bar{\alpha})\}. &
 \end{array}
 \]

$(iv) \Rightarrow (i)$ \\

Suppose now that $\xi^*$ satisfies the hypothesis, then
\[\max\limits_{x \in \mathcal{X}} F_{\phi}\{M(\xi^*, \bar{\beta}, \bar{\alpha}), m(x, \bar{\beta}, \bar{\alpha}))\} = 0\],
that means that $F_{\phi}\{ M(\xi^*, \bar{\beta},\bar{\alpha}), m(x,\bar{\beta},\bar{\alpha})\} \leq 0$, $\forall x \in \mathcal{X}$ .
According to the definition of the matrices $M \in \mathcal{M}$, any $M$ can be written as $M(\xi,\bar{\beta},\bar{\alpha} ) = \sum\limits_{i=1}^{r}\lambda_i m(x_i,\bar{\beta},\bar{\alpha} )$,
where $\sum\limits_{i=1}^{r}\lambda_i = 1$ and $\lambda_i > 0$ for every $i = 1, \ldots,r$.

Then, since $\phi$ is differentiable at $M(\xi, \bar{\beta},\bar{\alpha})$, it holds that:
\[F_{\phi}\{ M(\xi^*, \bar{\beta},\bar{\alpha} ), M(\xi,\bar{\beta},\bar{\alpha} )\} = \sum\limits_{i=1}^r \lambda_i F_{\phi}\{M(\xi^*, \bar{\beta},\bar{\alpha} ), m(x,\bar{\beta},\bar{\alpha} ) \} \leq 0\]
for every $\xi \in \Xi$.

This means, clearly, that

\[\phi(M(\xi,\bar{\beta},\bar{\alpha} )) - \phi(M(\xi^*,\bar{\beta},\bar{\alpha} )) \leq 0\]
 for every $\xi \in \Xi$, then $\xi^*$ is $\phi$-optimal.

\end{proof}

\end{theorem}

\subsection*{D-optimality}

Let consider now as design criterion the following function:
\[
\phi(M) = \left\{ \begin{array}{cc}
\log \det M & \text{if $M$ is non-singular} \\
-\infty & \text{otherwise} \end{array}\right.
\]
A design that maximizes such a $\phi$ function is called \emph{D-optimal design}.

In the case of D-optimality the Fr\'echet and the G\^ateaux derivatives have the following expression: \\

\emph{G\^ateaux derivative}
\[
\begin{array}{ccc}
\log\det(M_1 + \varepsilon M_2) - \log\det M_1 & = &\log\det(I + \varepsilon M_2 M_1^{-1}) = \\
= \log\{1 + \varepsilon \quad tr(M_2 M_1^{-1})\} + O(\varepsilon^2) & = & \varepsilon \quad tr(M_2M_1^{-1}) + O(\varepsilon^2)
\end{array}
\]
Hence, $G_{\phi}(M_1,M_2) = tr(M_2 M_1^{-1})$. \\

\emph{Fr\'echet derivative}
\[
F_{\phi}(M_1,M_2) =  G_{\phi}(M_1, M_2 - M_1)  =  tr((M_2 - M_1) M_1^{-1})= \]
\[
 = tr(M_2 M_1^{-1} - I_{(k+l)})  =  tr(M_2 M_1^{-1}) - (k+l)
\]
where $(k + l)$ is the number of the model parameters.

We are ready now to give an equivalence theorem which holds in the particular case of the D-criterion.

\begin{theorem}
For a fixed parameters vector $(\bar{\beta}, \bar{\alpha})$, the following properties are equivalent:
\begin{itemize}
\item $\xi^*$ is D-optimal;
\item $tr( M(\xi^*, \bar{\beta},\bar{\alpha})^{-1} m(x, \bar{\beta},\bar{\alpha}))\leq (k+l)$, $\forall x \in \mathcal{X}$;
\item $\xi^*$ minimize $\max\limits_{x \in \mathcal{X}}tr(M(\xi^*, \bar{\beta},\bar{\alpha})^{-1} m(x, \bar{\beta},\bar{\alpha}))$, over all $\xi \in \Xi$.
\end{itemize}

\begin{proof}
The proof comes directly from the Theorem \ref{Teo:EqT} by imputing the Fr\'echet derivative for the D-criterion.
\end{proof}

\end{theorem}

\section*{Bibliography}

\bibliographystyle{elsarticle-harv}
\bibliography{renreff-14perrone}


\end{document}